\documentclass[conference]{IEEEtran}
\IEEEoverridecommandlockouts

\usepackage{amsmath,amssymb,amsfonts}
\usepackage{bm}
\usepackage{graphicx}
\usepackage{booktabs}
\usepackage{tikz}
\usepackage{pgfplots}
\pgfplotsset{compat=1.17}
\usetikzlibrary{arrows.meta,positioning,fit,backgrounds}
\usepackage{xcolor}
\usepackage{cite}
\usepackage{bookmark}
\usepackage{algorithmic}
\usepackage{algorithm}

\definecolor{cmfmu}{HTML}{D62728}   
\definecolor{cmfno}{HTML}{F0997B}   
\definecolor{cdfmu}{HTML}{1F77B4}   
\definecolor{cdfno}{HTML}{85B7EB}   
\definecolor{cclstm}{HTML}{2CA02C}  

\def\BibTeX{{\rm B\kern-.05em{\sc i\kern-.025em b}\kern-.08em
    T\kern-.1667em\lower.7ex\hbox{E}\kern-.125emX}}

\begin{document}

\title{One-Step Generative CSI Prediction}

\author{
    \IEEEauthorblockN{Mehdi Sattari, Javad Aliakbari, Alexandre Graell i Amat, and Tommy Svensson}
    \IEEEauthorblockA{
        Department of Electrical Engineering, Chalmers University of Technology, Gothenburg, Sweden\\
        Email: \{mehdi.sattari, javada, alexandre.graell, tommy.svensson\}@chalmers.se
    }
    \thanks{
        This work was supported in part by the Swedish Research Council (VR) through the 6G-NTN-E Research Environment under Grant 2024-06645.
        This work was also partially supported by the Swedish Research Council (VR) under Grant 2023-05065; in part by the Wallenberg Artificial Intelligence (AI), Autonomous Systems and Software Program (WASP) funded by the Knut and Alice Wallenberg Foundation.
        The computations were enabled by resources provided by the National Academic Infrastructure for Supercomputing in Sweden (NAISS), 
        partially funded by the Swedish Research Council through grant agreement No.~2022-06725.
    }
}

\maketitle

\maketitle

\begin{abstract}
Channel state information (CSI) prediction has emerged as a promising approach for mitigating channel aging in wireless communication systems. Recently, generative models have been employed to develop CSI prediction frameworks capable of modeling the uncertainty in future channel realizations. However, the high computational complexity and inference latency of these models remain major obstacles to their practical deployment. To address these challenges, this paper focuses on a class of one-step generative models, namely MeanFlow models, in which inference requires only a single neural function evaluation (NFE), significantly reducing computational cost and latency. To alleviate the difficulty of generating CSI directly from noise, we introduce a channel-informed prior that initializes the generation process from an informative estimate of the future channel rather than pure Gaussian noise. Incorporating this channel-informed prior substantially improves both point prediction accuracy and uncertainty calibration. Furthermore, we compare the proposed MeanFlow-based CSI prediction framework with a recent diffusion-based approach. The results demonstrate that MeanFlow achieves comparable point prediction accuracy while producing better-calibrated CSI predictions with only a single NFE.
\end{abstract}


\section{Introduction}
Massive multiple-input multiple-output (MIMO) systems rely on accurate channel state information (CSI) for precoding and link adaptation. However, channel aging and pilot overhead remain major bottlenecks that limit the efficient utilization of wireless channels \cite{6736761}. CSI prediction, which forecasts future CSI from past observations, has emerged as a promising approach to mitigate these challenges. Classical prediction techniques, including Kalman filtering \cite{9210016}, Prony's method \cite{9127447}, and autoregressive (AR) models \cite{1512123}, have been extensively studied. However, these methods typically rely on restrictive and often unrealistic assumptions about channel dynamics, limiting their performance in practical wireless environments.

More recently, deep learning models have been proposed to approximate the optimal predictor directly from data, resulting in substantial improvements in CSI prediction accuracy \cite{9044427, 9832933, 10965849, 10978424}. However, most existing approaches are deterministic and produce only a single estimate of the future channel. They therefore provide no direct measure of predictive uncertainty, making it difficult to assess prediction reliability and support risk-aware link adaptation. Additionally, they struggle to capture the stochastic and potentially multimodal nature of channel evolution~\cite{11644834, 11587923}.

Generative models address these limitations by learning the full conditional distribution of future CSI. Among them, diffusion-based CSI prediction frameworks have demonstrated strong predictive performance~\cite{NEURIPS2020_4c5bcfec,11644834, 11587923}. However, their inference procedure relies on an iterative multi-step reverse denoising process, resulting in high computational complexity and inference latency. Flow matching and its one-step variant, MeanFlow~\cite{lipman2023flowmatchinggenerativemodeling,geng2025meanflowsonestepgenerative}, offer efficient alternatives for generative modeling. By learning an average velocity field, MeanFlow generates samples using only a single neural function evaluation (NFE), making it particularly attractive for real-time CSI prediction.

Despite its low inference complexity, MeanFlow has not yet been investigated
for CSI prediction. Motivated by its potential for practical, low-latency
inference, we propose a one-step generative CSI prediction framework based on MeanFlow. Conventional MeanFlow initializes the generation process with
zero-mean white Gaussian noise. However, this uninformative source distribution does not reflect the statistical or temporal structure of CSI and can be far from the distribution of future CSI, making an accurate one-step transformation challenging. We therefore adapt MeanFlow to CSI prediction by introducing a channel-informed source distribution centered on a coarse point estimate of the future CSI.

We compare the proposed MeanFlow framework with the diffusion-based CSI prediction scheme in \cite{11644834} and evaluate both point prediction accuracy, measured by the normalized mean squared error (NMSE), and predictive uncertainty using calibration metrics, including prediction interval coverage probability (PICP) and continuous ranked probability score (CRPS).
Our results demonstrate that incorporating the informative prior significantly improves the point prediction accuracy of both diffusion- and MeanFlow-based generators. Furthermore, despite requiring only a single NFE, the MeanFlow generator produces substantially better-calibrated predictive distributions than the multi-step diffusion generator while maintaining comparable point prediction accuracy. We further show that point prediction accuracy can be improved beyond that of multi-step diffusion by increasing the capacity of the MeanFlow generator, with only a modest increase in inference latency. Finally, we demonstrate that MeanFlow's improved calibration translates into more accurate achievable-rate prediction, as measured by CRPS-rate.

\section{Problem Formulation and Preliminaries}
\subsection{Problem Formulation}
We consider a MIMO system in which a base station (BS) equipped with a uniform linear array (ULA) of \(N_t\) antennas serves single-antenna users over \(N_c\) orthogonal frequency division multiplexing (OFDM) subcarriers. The spatial--frequency CSI at time index \(n\) is denoted by \(\mathbf{H}_n \in \mathbb{C}^{N_t \times N_c}\). Given a sequence of past CSI observations,
\begin{equation}
\mathbf{H}_{\mathrm{p}}
=
\left\{
\mathbf{H}_{n-N_p+1},
\ldots,
\mathbf{H}_n
\right\},
\end{equation}
the objective is to predict the future CSI sequence
\begin{equation}
\mathbf{H}_{\mathrm{f}}
=
\left\{
\mathbf{H}_{n+1},
\ldots,
\mathbf{H}_{n+N_f}
\right\}.
\end{equation}

Under the mean squared error (MSE) criterion, the optimal predictor is the conditional mean,
\begin{equation}
\label{eq:CME}
\begin{aligned}
f^\ast(\mathbf{H}_{\mathrm{p}})
&=
\mathbb{E}
\!\left[
\mathbf{H}_{\mathrm{f}}
\mid
\mathbf{H}_{\mathrm{p}}
\right] \\
&=
\int
\mathbf{H}_{\mathrm{f}}\,
p(\mathbf{H}_{\mathrm{f}}
\mid
\mathbf{H}_{\mathrm{p}})
\,d\mathbf{H}_{\mathrm{f}}.
\end{aligned}
\end{equation}

Equivalently, the conditional mean is the minimum mean squared error (MMSE) predictor,
\begin{equation}
f^\ast(\mathbf{H}_{\mathrm{p}})
=
\arg\min_f
\mathbb{E}
\!\left[
\left\|
\mathbf{H}_{\mathrm{f}}
-
f(\mathbf{H}_{\mathrm{p}})
\right\|^2
\right].
\end{equation}

Computing the MMSE predictor is generally intractable because the conditional distribution
\(p(\mathbf{H}_{\mathrm{f}} \mid \mathbf{H}_{\mathrm{p}})\)
is unknown and the underlying wireless channel dynamics are highly nonlinear and stochastic. Consequently, data-driven methods, particularly deep learning models, have been widely adopted to approximate the conditional mean directly from data.

\subsection{MeanFlow Models}

MeanFlow~\cite{geng2025meanflowsonestepgenerative} is a flow-matching-based generative model that directly estimates the average transport velocity between a source distribution and the target data distribution. Unlike diffusion models or conventional flow matching, which require multiple numerical integration steps during inference, MeanFlow generates samples with a single network evaluation.

Given a data sample $\mathbf{Y}$ and a source sample $\mathbf{S}$, MeanFlow defines the interpolation
\begin{equation}\label{eq:MF_interpolation}
\mathbf{H}^{t}=(1-t)\mathbf{Y}+t\mathbf{S},
\qquad
t\in[0,1],
\end{equation}
where $t$ denotes the interpolation time. During training, an additional time variable $r\in[0,t]$ is sampled to define the averaging interval $[r,t]$ over which the transport velocity is estimated. Instead of learning the instantaneous velocity $\mathbf{v}=\mathbf{S}-\mathbf{Y}$, MeanFlow learns the average transport velocity
\begin{equation}
\mathbf{u}_{\theta}(\mathbf{H}^{t},r,t),
\end{equation}
which is trained using the objective proposed in~\cite{geng2025meanflowsonestepgenerative}. Once trained, the prediction is obtained by
\begin{equation}
\hat{\mathbf{Y}}
=
\mathbf{S}
-
\mathbf{u}_{\theta},
\end{equation}
requiring only a single network evaluation.

\section{One-Step Generative CSI Prediction}\label{sec:methods}

\begin{figure}[t]
\centering
\resizebox{\columnwidth}{!}{%
\begin{tikzpicture}[
  font=\footnotesize,
  box/.style={draw, rounded corners, align=center, text width=23mm, minimum height=11mm, inner sep=2pt},
  io/.style={draw, align=center, text width=13mm, minimum height=9mm, inner sep=2pt, fill=black!5},
  ar/.style={-{Latex[length=1.8mm]}, semithick},
]
  \node[io] (hp) {$\mathbf{H}_p$};
  \node[box, fill=cdfmu!12, right=9mm of hp] (enc) {Temporal Encoder};
  \node[box, fill=cmfmu!12, right=27mm of enc] (gen) {MeanFlow Generator};
  \node[io, right=10mm of gen] (out) {$\hat{\mathbf{H}}_{n+1}$};
  \node[align=center, above=8mm of gen, font=\scriptsize] (src) {source\\$\mathbf{S}=\bm{\mu}+\sigma\bm{\epsilon}$};

  \draw[ar] (hp) -- (enc);
  \draw[ar] (enc) -- node[above, font=\scriptsize]{$\mathbf{z}$} (gen);
  \draw[ar] (enc.north) |- (src.west) node[pos=0.28, left, font=\scriptsize]{$\bm{\mu}$};
  \draw[ar] (src) -- (gen);
  \draw[ar] (gen) -- (out);
  \draw[ar, dashed] (out.south) -- ++(0,-1.3) -| (hp.south);
\end{tikzpicture}%
}


\caption{One-step generative CSI predictor using MeanFlow generator.}
\label{fig:arch}
\end{figure}
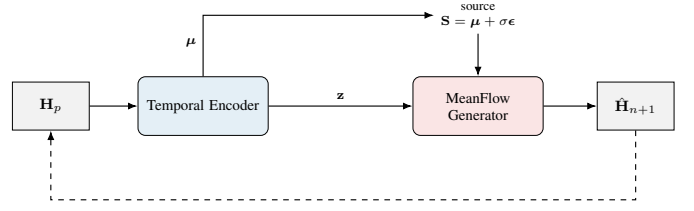

Fig.~\ref{fig:arch} illustrates the overall framework of the proposed one-step generative CSI prediction method. While MeanFlow was originally proposed for image generation, we adapt it to conditional CSI prediction by conditioning the generative process on temporal features extracted from historical CSI. The framework consists of a temporal encoder and a MeanFlow generator. The temporal encoder learns latent representations that capture the channel dynamics, and the MeanFlow generator uses these representations to generate future CSI in a single network evaluation.

\subsection{MeanFlow Training}

Given the historical CSI sequence $\mathbf{H}_{\mathrm{p}}$, the temporal encoder extracts a latent representation $\mathbf{z}$ together with a deterministic prediction
\begin{equation}
\boldsymbol{\mu}\approx\mathbb{E}[\mathbf{Y}\mid\mathbf{H}_{\mathrm{p}}],
\end{equation}
where $\mathbf{Y}$ denotes the ground-truth future CSI. The latent representation $\mathbf{z}$ captures the temporal channel dynamics and conditions the MeanFlow generator, while $\boldsymbol{\mu}$ provides an informative prior for the source distribution.

Instead of sampling the source point from a zero-mean Gaussian distribution as in the original MeanFlow formulation, we center the source around the encoder prediction,
\begin{equation}
\mathbf{S}
=
\operatorname{sg}(\boldsymbol{\mu})
+
\boldsymbol{\epsilon},
\qquad
\boldsymbol{\epsilon}\sim\mathcal{N}(\mathbf{0},\sigma^2\mathbf{I}),
\end{equation}
where $\operatorname{sg}(\cdot)$ denotes the stop-gradient operator.
By centering the source on $\boldsymbol{\mu}$, the predictable component is
provided directly to the generator, allowing it to focus primarily on the
remaining prediction residual. This places the source distribution closer to
the conditional distribution of the future CSI and reduces the transport
distance that MeanFlow must learn, which is particularly important for
one-step generation. Meanwhile, the Gaussian perturbation
$\boldsymbol{\epsilon}$ preserves the stochasticity required to represent
multiple possible future channel realizations.

The interpolated state defined in \eqref{eq:MF_interpolation} is provided to the MeanFlow generator together with the conditioning latent representation $\mathbf{z}$. The target average velocity is computed using the MeanFlow identity and a Jacobian--vector product (JVP), following the original MeanFlow formulation. As in~\cite{geng2025meanflowsonestepgenerative}, the target velocity is detached from gradient propagation, and adaptive loss weighting is employed to improve optimization stability. The complete training procedure is summarized in Algorithm~\ref{alg:meanflow_training_procedure}. In addition to the MeanFlow objective, the temporal encoder is optimized using an auxiliary MSE loss on the deterministic prediction $\boldsymbol{\mu}$.

\begin{algorithm}[t]
\caption{MeanFlow Training}
\label{alg:meanflow_training_procedure}
\textbf{Input:}
Dataset $\mathcal{D} = \{(\mathbf{X}, \mathbf{Y})\}$;
parameters $\boldsymbol{\theta} = [\boldsymbol{\theta}_{\mathrm{TE}}, \boldsymbol{\theta}_{\mathrm{G}}]$;
source scale $\sigma$; loss weights $\lambda, c, p$;
SNR range $[\rho_{\min}, \rho_{\max}]$.
\\
\textbf{Output:} $\boldsymbol{\theta}^\ast$.
\begin{algorithmic}[1]
    \FOR{$\mathrm{epoch} = 1, \dots, N_{\mathrm{epochs}}$}
        \FOR{each batch $(\mathbf{X}, \mathbf{Y}) \in \mathcal{D}$}
            \STATE $\rho \sim \mathcal{U}[\rho_{\min}, \rho_{\max}]$ 
            \STATE $\tilde{\mathbf{X}} \gets \mathbf{X} + \sigma_{\rho}\mathbf{N}$,\quad
                   $\mathbf{N} \sim \mathcal{N}(\mathbf{0}, \mathbf{I})$,\quad
                   $\sigma_{\rho}^2 = P_{\mathbf{X}}/10^{\rho/10}$
            \STATE $(\mathbf{Z}, \boldsymbol{\mu}) \gets f_{\boldsymbol{\theta}_{\mathrm{TE}}}(\tilde{\mathbf{X}})$
            \STATE Sample $t, r \in [0,1]$ with $t \geq r$
            \STATE $\boldsymbol{\epsilon} \sim \mathcal{N}(\mathbf{0}, \sigma^2\mathbf{I})$
            \STATE $\mathbf{S} \gets \mathrm{sg}(\boldsymbol{\mu}) + \boldsymbol{\epsilon}$
            \STATE $\mathbf{H}^{t} \gets (1-t)\mathbf{Y} + t\mathbf{S}$,\quad
                   $\mathbf{v} \gets \mathbf{S} - \mathbf{Y}$
            \STATE $\mathbf{u}_{\boldsymbol{\theta}} \gets
                   f_{\boldsymbol{\theta}_{\mathrm{G}}}(\mathrm{Concat}(\mathbf{H}^{t}, \mathbf{Z}), r, t)$
            \STATE $\dfrac{d\mathbf{u}_{\boldsymbol{\theta}}}{dt} \gets
                   \mathbf{v}\,\partial_{\mathbf{H}^{t}}\mathbf{u}_{\boldsymbol{\theta}}
                   + \partial_t \mathbf{u}_{\boldsymbol{\theta}}$
            \STATE $\mathbf{u}_{\mathrm{tgt}} \gets
                   \mathbf{v} - (t-r)\dfrac{d\mathbf{u}_{\boldsymbol{\theta}}}{dt}$
            \STATE $w \gets \big(\|\mathbf{u}_{\boldsymbol{\theta}} - \mathbf{u}_{\mathrm{tgt}}\|_2^2 + c\big)^{-p}$
            \STATE $\mathcal{L}(\boldsymbol{\theta}) \gets
                   \mathrm{sg}(w)\,\big\|\mathbf{u}_{\boldsymbol{\theta}} - \mathrm{sg}(\mathbf{u}_{\mathrm{tgt}})\big\|_2^2
                   + \lambda\,\|\boldsymbol{\mu} - \mathbf{Y}\|_2^2$
            \STATE Update $\boldsymbol{\theta}$ using $\nabla_{\boldsymbol{\theta}}\mathcal{L}(\boldsymbol{\theta})$
        \ENDFOR
    \ENDFOR
    \STATE \textbf{Return} $\boldsymbol{\theta}^\ast$
\end{algorithmic}
\end{algorithm}

\subsection{One-Step AR Inference}

During inference, the temporal encoder first processes the available CSI history to produce the conditioning latent $\mathbf{z}$ and the deterministic prediction $\boldsymbol{\mu}$. A source sample is then generated by perturbing the informative prior,
\begin{equation}
\mathbf{H}^{1}
=
\boldsymbol{\mu}
+
\sigma\boldsymbol{\epsilon},
\qquad
\boldsymbol{\epsilon}\sim\mathcal{N}(\mathbf{0},\mathbf{I}).
\end{equation}

Unlike diffusion models or conventional flow matching, no iterative denoising or ordinary differential equation (ODE) integration is required. The MeanFlow generator is evaluated only once to estimate the average transport velocity between the source and target distributions,
\begin{equation}
\mathbf{u}_{\boldsymbol{\theta}}
=
f_{\boldsymbol{\theta}_{\mathrm{G}}}
\bigl(\mathrm{Concat}(\mathbf{H}^{1},\mathbf{z}),0,1\bigr),
\end{equation}
and the next CSI frame is recovered by
\begin{equation}
\hat{\mathbf{H}}_{n}
=
\mathbf{H}^{1}
-
\mathbf{u}_{\boldsymbol{\theta}}.
\end{equation}

The predicted frame is appended to the history and used as input for the next prediction step. This AR procedure is repeated until the desired prediction horizon is reached, as summarized in Algorithm~\ref{alg:meanflow_inference_AR}. Since each prediction requires only a single evaluation of the MeanFlow generator, the proposed method requires 1-NFE per predicted CSI frame, resulting in substantially lower inference complexity than iterative generative models.

\begin{algorithm}[t]
\caption{MeanFlow AR Inference}
\label{alg:meanflow_inference_AR}
\textbf{Input:} History $\mathbf{H}_{\mathrm{p}}$;
trained $\boldsymbol{\theta}^{\ast} = [\boldsymbol{\theta}_{\mathrm{TE}}^{\ast}, \boldsymbol{\theta}_{\mathrm{G}}^{\ast}]$;
horizon $N_{\mathrm{f}}$; source scale $\sigma$.
\\
\textbf{Output:}
$\{\hat{\mathbf{H}}_n\}_{n=1}^{N_{\mathrm{f}}}$.
\begin{algorithmic}[1]
    \STATE $\mathbf{H}_{c} \gets \mathbf{H}_{\mathrm{p}}$
    \FOR{$n = 1$ \TO $N_{\mathrm{f}}$}
        \STATE $(\mathbf{Z}, \boldsymbol{\mu}) \gets f_{\boldsymbol{\theta}^{\ast}_{\mathrm{TE}}}(\mathbf{H}_{c})$
        \STATE $\mathbf{H}^{1} \gets \boldsymbol{\mu} + \sigma\boldsymbol{\epsilon}$,\quad
               $\boldsymbol{\epsilon} \sim \mathcal{N}(\mathbf{0}, \mathbf{I})$
        \STATE $\mathbf{u}_{\boldsymbol{\theta}^{\ast}} \gets
               f_{\boldsymbol{\theta}^{\ast}_{\mathrm{G}}}(\mathrm{Concat}(\mathbf{H}^{1}, \mathbf{Z}), 0, 1)$
        \STATE $\hat{\mathbf{H}}_n \gets \mathbf{H}^{1} - \mathbf{u}_{\boldsymbol{\theta}^{\ast}}$
        \STATE $\mathbf{H}_{c} \gets \mathrm{Concat}(\mathbf{H}_{c}, \hat{\mathbf{H}}_n)$
    \ENDFOR
    \STATE \textbf{Return} $\{\hat{\mathbf{H}}_n\}_{n=1}^{N_{\mathrm{f}}}$
\end{algorithmic}
\end{algorithm}

\section{Simulation Results}\label{sec:results}

\subsection{Dataset Setup}
We generate CSI on the fly using the 3GPP CDL channel model using Sionna channel generator \cite{hoydis2023sionnaopensourcelibrarynextgeneration}. Each sample is a tensor of shape
$[T \times 2 \times N_\mathrm{t} \times N_\mathrm{c}]$, where $T = N_\mathrm{p}+N_\mathrm{f}$
is the number of OFDM symbols, followed by real/imaginary parts, $N_\mathrm{t}$ transmit
antennas, and $N_\mathrm{c}$ subcarriers. Each sample is split into past CSI
$\mathbf{X}=\mathbf{H}_\mathrm{p}$ ($N_\mathrm{p}=30$) and future CSI
$\mathbf{Y}=\mathbf{H}_\mathrm{f}$ ($N_\mathrm{f}=10$), giving $T=40$. Samples are
standardized to zero mean and unit variance. Table~\ref{tab:simulation_setup} summarizes
the dataset parameters.

\begin{table}[t]
\centering
\caption{Dataset Parameters}
\label{tab:simulation_setup}
\renewcommand{\arraystretch}{1.2}
\begin{tabular}{ll}
\toprule
\textbf{Parameter} & \textbf{Value} \\
\midrule
Carrier frequency $f_c$ & 28 GHz \\
Channel model & 3GPP \texttt{CDL-A}–\texttt{E} (randomly selected) \\
Number of BS antennas $N_\mathrm{t}$ & 16 \\
Resource blocks & 25 (300 subcarriers) \\
Subcarrier spacing & 30 kHz \\
Used subcarriers $N_\mathrm{c}$ & 16 (evenly spaced) \\
Symbol duration & $\approx 33.3\,\mu$s \\
User velocity & 30–120 km/h (uniform) \\
Delay spread & 50–400 ns (uniform) \\
Past / future length & $N_\mathrm{p}=30$ / $N_\mathrm{f}=10$ \\
\bottomrule
\end{tabular}
\end{table}

\subsection{Network Architecture and Training}
To ensure a controlled comparison, the diffusion and MeanFlow models share the same
backbone and differ only in their training objective. A single-layer ConvLSTM temporal
encoder ($128$ hidden channels) maps the history to a $128$-channel conditioning latent
$\mathbf{Z}$ and a $2$-channel point estimate $\boldsymbol{\mu}$. The generator is a
2D U-Net (base width $32$, one downsampling stage to $64$ channels, two residual blocks
per stage, single-head self-attention at the bottleneck) that takes the noisy CSI frame
concatenated with $\mathbf{Z}$ and is conditioned on the time pair $(r,t)$ or the
diffusion step $t$ through a sinusoidal FiLM embedding. Diffusion predicts the clean
frame under a cosine noise schedule and samples with $3$-step DDIM, whereas MeanFlow
learns the average velocity and samples in a single NFE. The encoder and generator
together comprise ${\approx}2.2$\,M parameters.

All models are trained identically with Adam (learning rate $2{\times}10^{-4}$, cosine
schedule with warmup), batch size $256$, for $50$k steps, with gradient clipping and an
exponential moving average of the weights used at evaluation. The history is augmented
with additive noise at a random SNR to emulate imperfect CSI. Full model and training
configurations are available in our GitHub implementation.\footnote{\url{https://github.com/MehdiSattari/meanflow-csi-prediction}}

\subsection{Evaluation Metrics}
We assess both the accuracy of the point prediction and the quality of the predictive
distribution obtained from $K{=}30$ samples
$\{\hat{\mathbf{H}}^{(k)}\}_{k=1}^{K}$. All metrics are computed in the physical
(denormalized) channel space. For a single real-valued coefficient, let $\{x_k\}_{k=1}^{K}$
be the ensemble values, $x_{(1)}\!\le\!\cdots\!\le\! x_{(K)}$ their order statistics,
$\bar{x}=\tfrac1K\sum_k x_k$ the ensemble mean, and $y$ the corresponding ground-truth
value; expectations $\mathbb{E}[\cdot]$ denote averaging over coefficients (antennas, subcarriers, and real/imaginary parts) and the batch.
For evaluations, we consider the following metrics:
\begin{itemize}

\item \textbf{NMSE.}
NMSE is computed using the ensemble mean
$\bar{\mathbf{H}}=\frac{1}{K}\sum_{k=1}^{K}\hat{\mathbf{H}}^{(k)}$,
\begin{equation}
\mathrm{NMSE}_n =
\frac{\mathbb{E}\!\left[\|\bar{\mathbf{H}}_n-\mathbf{H}_n\|_2^2\right]}
     {\mathbb{E}\!\left[\|\mathbf{H}_n\|_2^2\right]} .
\end{equation}

\item \textbf{PICP.}
PICP measures how often the
true value falls within a prediction interval. For a nominal confidence level
$p$, it is defined as
\begin{equation}
    \mathrm{PICP}(p)
    =
    \mathbb{E}\!\left[
    \mathbf{1}\!\left\{
    q_{\frac{1-p}{2}}
    \le y \le
    q_{\frac{1+p}{2}}
    \right\}
    \right],
\end{equation}
where $q_\alpha$ is the empirical $\alpha$-quantile of the generated samples.
A well-calibrated model should satisfy $\mathrm{PICP}(p)\approx p$.

\item \textbf{CRPS.}
CRPS evaluates the quality of the
entire predictive distribution by considering both accuracy and sharpness.
For a predictive cumulative distribution function (CDF) $F$ and a true value
$y$, it is defined as
\begin{equation}
    \mathrm{CRPS}(F,y)
    =
    \int_{-\infty}^{\infty}
    \left(
    F(x)-\mathbf{1}\{x\geq y\}
    \right)^2
    \,\mathrm{d}x .
\end{equation}
A lower CRPS indicates a better probabilistic prediction.

\item \textbf{Achievable-rate CRPS.}
To assess predictive uncertainty from a communication perspective, we evaluate
the predicted distribution of the maximum-ratio achievable rate. For a channel
realization $\mathbf{h}$ and operating SNR $\gamma$, the rate is
\begin{equation}
    C(\mathbf{h})
    =
    \log_2\!\left(1+\gamma\|\mathbf{h}\|_2^2\right).
\end{equation}
Each generated CSI sample $\hat{\mathbf{h}}^{(k)}$ is mapped to its
corresponding rate $C(\hat{\mathbf{h}}^{(k)})$. The achievable-rate CRPS is
then computed by comparing the resulting rate distribution with the true rate
$C(\mathbf{h})$.
\end{itemize}

All metrics are reported per prediction step and as horizon averages.

\subsection{Numerical Results}
\begin{figure}[t]
\centering
\begin{tikzpicture}
\begin{axis}[width=\columnwidth,height=6.0cm,
  xlabel={prediction step}, ylabel={NMSE (dB)},
  xmin=1,xmax=10, grid=both, grid style={gray!20},
  legend style={
    font=\scriptsize,
    at={(0.98,0.02)},
    anchor=south east,
    draw=none,
    fill=white,
    fill opacity=0.8
    },
  legend cell align=left, tick label style={font=\scriptsize},
  label style={font=\small}]
\addplot[cmfmu,mark=*,thick] coordinates{(1,-23.10)(2,-20.35)(3,-17.65)(4,-15.24)(5,-13.14)(6,-11.34)(7,-9.78)(8,-8.39)(9,-7.15)(10,-6.01)};
\addplot[cdfmu,mark=square*,thick] coordinates{(1,-22.05)(2,-19.30)(3,-16.77)(4,-14.59)(5,-12.78)(6,-11.26)(7,-9.97)(8,-8.83)(9,-7.83)(10,-6.93)};
\addplot[cmfno,mark=o,thick] coordinates{(1,-18.93)(2,-15.43)(3,-12.45)(4,-10.09)(5,-8.23)(6,-6.77)(7,-5.57)(8,-4.56)(9,-3.69)(10,-2.92)};
\addplot[cdfno,mark=square,thick] coordinates{(1,-10.13)(2,-9.45)(3,-7.78)(4,-6.15)(5,-4.92)(6,-4.06)(7,-3.41)(8,-2.85)(9,-2.39)(10,-1.91)};
\legend{$\mu$-MeanFlow,$\mu$-Diffusion,MeanFlow,Diffusion}
\end{axis}
\end{tikzpicture}
\vspace{-4mm}
\caption{Per-step NMSE at an inference SNR of $20$\,dB.}
\label{fig:nmse}
\end{figure}
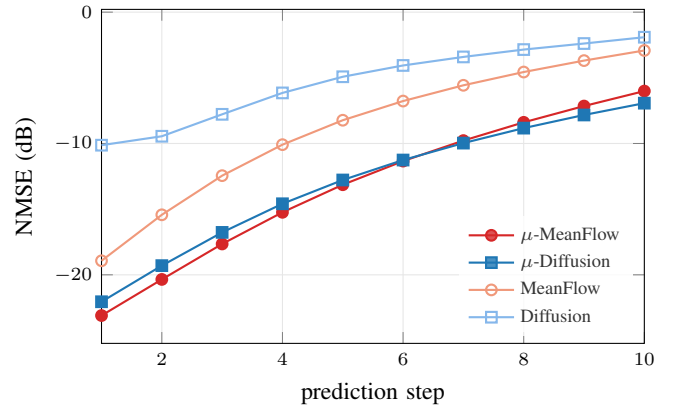

Fig.~\ref{fig:nmse} shows the NMSE versus the prediction horizon. For both
Diffusion and MeanFlow, we compare the conventional Gaussian source
distribution with the channel-informed prior, denoted by $\mu$-Diffusion and
$\mu$-MeanFlow. Incorporating the
informative prior improves the average NMSE of both models by approximately
$5$~dB, with $\mu$-Diffusion and $\mu$-MeanFlow achieving nearly identical
performance. This result confirms that centering the MeanFlow source
distribution or the diffusion target around $\bm{\mu}$ replaces the
challenging task of generating CSI directly from noise with the simpler task
of modeling a zero-mean residual. Without the informative prior $\bm{\mu}$, MeanFlow achieves better NMSE than Diffusion despite requiring fewer NFEs.

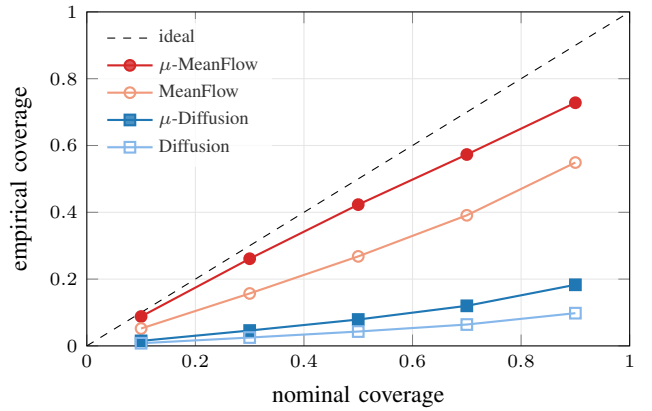
\begin{figure}[t]
\centering
\begin{tikzpicture}
\begin{axis}[width=\columnwidth,height=6.0cm,
  xlabel={nominal coverage}, ylabel={empirical coverage},
  xmin=0,xmax=1,ymin=0,ymax=1, grid=both, grid style={gray!20},
  legend style={font=\scriptsize,at={(0.02,0.98)},anchor=north west,draw=none,fill=white,fill opacity=0.8},
  legend cell align=left, tick label style={font=\scriptsize}, label style={font=\small}]
\addplot[black,dashed] coordinates{(0,0)(1,1)};
\addplot[cmfmu,mark=*,thick] coordinates{(0.1,0.088)(0.3,0.261)(0.5,0.423)(0.7,0.573)(0.9,0.728)};
\addplot[cmfno,mark=o,thick] coordinates{(0.1,0.052)(0.3,0.157)(0.5,0.268)(0.7,0.391)(0.9,0.549)};
\addplot[cdfmu,mark=square*,thick] coordinates{(0.1,0.015)(0.3,0.046)(0.5,0.079)(0.7,0.120)(0.9,0.183)};
\addplot[cdfno,mark=square,thick] coordinates{(0.1,0.008)(0.3,0.025)(0.5,0.043)(0.7,0.064)(0.9,0.098)};
\legend{ideal,$\mu$-MeanFlow ,MeanFlow,$\mu$-Diffusion,Diffusion}
\end{axis}
\end{tikzpicture}
\vspace{0mm}
\caption{Reliability at an inference SNR of $20$\,dB.}
\label{fig:reliability}
\end{figure}

Figure~\ref{fig:reliability} compares the calibration of the predictive distributions using reliability diagrams. MeanFlow closely follows the ideal diagonal, achieving a coverage of $0.73$ at the nominal $0.90$ confidence level, whereas diffusion is overconfident with a coverage of $0.18$, indicating that the true CSI frequently lies outside the predicted ensemble. Consistent with this observation, Table~\ref{tab:summary} shows that MeanFlow also achieves the lowest CRPS while maintaining nearly identical NMSE to diffusion. Moreover, this calibration advantage persists both with and without the informative prior, suggesting that it primarily originates from the flow-matching formulation rather than the informative prior itself.

\begin{table}[t]
\centering
\caption{Prediction accuracy and calibration at inference SNRs of $20\,\textnormal{dB}$ and $5\,\textnormal{dB}$.}
\label{tab:summary}
\setlength{\tabcolsep}{4pt}
\begin{tabular}{lccccccc}
\toprule
& \multicolumn{3}{c}{$20$\,dB} & \multicolumn{3}{c}{$5$\,dB}\\
\cmidrule(lr){2-4}\cmidrule(lr){5-7}
Model & NMSE & CRPS & cov$_{90}$ & NMSE & CRPS & cov$_{90}$\\
\midrule
$\mu$-MeanFlow  & $-10.6$ & $\mathbf{0.10}$ & $\mathbf{0.73}$ & $-9.8$  & $\mathbf{0.12}$ & $\mathbf{0.59}$\\
$\mu$-Diffusion & $\mathbf{-11.0}$ & $0.11$ & $0.18$ & $\mathbf{-10.0}$ & $0.14$ & $0.31$\\
MeanFlow  & $-6.7$  & $0.17$ & $0.55$ & $-7.1$  & $0.17$ & $0.49$\\
Diffusion & $-4.5$  & $0.28$ & $0.10$ & $-5.5$  & $0.25$ & $0.19$\\
\bottomrule
\end{tabular}
\end{table}

\begin{table}[t]
\centering
\caption{Achievable-rate CRPS and $90\%$ coverage across various inference SNRs.}
\label{tab:snr}
\setlength{\tabcolsep}{6pt}
\begin{tabular}{lcccc}
\toprule
 & \multicolumn{4}{c}{Inference SNR (dB)}\\
\cmidrule(lr){2-5}
Model & $0$ & $5$ & $10$ & $20$\\
\midrule
\multicolumn{5}{l}{\textit{CRPS of achievable rate}}\\
$\mu$-MeanFlow & $\mathbf{0.16}$ & $\mathbf{0.14}$ & $\mathbf{0.13}$ & $0.17$\\
$\mu$-Diffusion & $0.21$ & $0.18$ & $0.16$ & $0.16$\\
MeanFlow  & $0.20$ & $0.21$ & $0.25$ & $0.30$\\
Diffusion & $0.76$ & $0.77$ & $0.76$ & $0.71$\\
\midrule
\multicolumn{5}{l}{\textit{Coverage @ $0.90$}}\\
$\mu$-MeanFlow  & $\mathbf{0.57}$ & $\mathbf{0.59}$ & $\mathbf{0.63}$ & $\mathbf{0.72}$\\
$\mu$-Diffusion & $0.36$ & $0.31$ & $0.24$ & $0.18$\\
MeanFlow  & $0.49$ & $0.49$ & $0.51$ & $0.55$\\
Diffusion & $0.24$ & $0.18$ & $0.13$ & $0.10$\\
\bottomrule
\end{tabular}
\end{table}

Next, Table~\ref{tab:snr} evaluates probabilistic predictions across inference SNRs. All models under-cover relative to the nominal $0.90$ level. $\mu$-MeanFlow achieves the closest coverage, ranging from $0.57$ at $0$\,dB to $0.72$ at $20$\,dB, although its intervals become less reliable as the input SNR decreases. Conversely, $\mu$-Diffusion coverage improves from $0.18$ at $20$\,dB to $0.36$ at $0$\,dB but remains far below the nominal level. $\mu$-MeanFlow also achieves the lowest achievable-rate CRPS at $0$--$10$\,dB. At $20$\,dB, $\mu$-Diffusion has a marginally lower CRPS ($0.16$ versus $0.17$) but much lower coverage ($0.18$ versus $0.72$). Overall, $\mu$-MeanFlow provides the best balance between CRPS and coverage.


To assess whether improved calibration translates into better communication performance, we report the achievable-rate CRPS in Table~\ref{tab:snr}. MeanFlow with prior achieves the lowest achievable-rate CRPS in the low- and moderate-SNR regime ($0$--$10$\,dB), outperforming few-step diffusion. At $20$\,dB, the performance of the two generative models becomes comparable because the achievable rate approaches saturation. These results demonstrate that the superior calibration of MeanFlow translates into measurable downstream performance gains precisely in the regime where predictive uncertainty is most pronounced.

\begin{table}[t]
\centering
\caption{Effect of the number of DDIM sampling steps on diffusion accuracy and reliability.}
\label{tab:diffsteps}
\begin{tabular}{lcc}
\toprule
Model (NFE/frame) & NMSE & cov$_{90}$ \\
\midrule
$\mu$-Diffusion ($3$)  & $-11.2$ & $0.19$ \\
$\mu$-Diffusion ($10$) & $-11.1$ & $0.37$ \\
$\mu$-Diffusion ($20$) & $-10.9$ & $0.47$ \\
$\mu$-Diffusion ($50$) & $-10.7$ & $0.54$ \\
\midrule
$\mu$-MeanFlow ($1$)   & $-10.5$ & $\mathbf{0.72}$ \\
\bottomrule
\end{tabular}
\end{table}

Finally, Table~\ref{tab:diffsteps} investigates whether diffusion's overconfidence is primarily caused by using only three DDIM sampling steps. 
Increasing the number of sampling steps consistently improves calibration, with the $90\%$ coverage increasing from $0.19$ to $0.54$. This improvement, however, comes at a slight cost in ensemble NMSE ($-11.2$ to $-10.7$\,dB), since a more faithful sampler yields a more diverse ensemble whose mean is a slightly noisier point estimate. Despite this trade-off, diffusion remains less calibrated than one-step $\mu$-MeanFlow even after $50$ DDIM steps, achieving a coverage of only $0.54$ versus $0.72$. These results indicate that increasing the number of DDIM steps alleviates, but does not eliminate, diffusion's overconfidence.


\subsection{Inference Complexity and Latency}
Since MeanFlow and diffusion share the same temporal encoder and generator, they have
identical parameter counts ($2.2$\,M); their inference costs differ only in the number of
generator evaluations (NFE). We analyze the cost of a single prediction step for a single
user, which is the latency-critical operating point for real-time CSI prediction. Because
the temporal encoder is a recurrent ConvLSTM, each autoregressive step advances its hidden
state by one frame.
The per-step cost is therefore
\begin{equation}
C_{\mathrm{step}} = C_{\mathrm{enc}} + N_{\mathrm{NFE}}\,C_{\mathrm{gen}},
\end{equation}
where $C_{\mathrm{enc}}$ is one recurrent encoder step and $C_{\mathrm{gen}}$ is one
generator (U-Net) evaluation; MeanFlow uses $N_{\mathrm{NFE}}{=}1$, while diffusion uses
$N_{\mathrm{NFE}}\ge 3$.

Table~\ref{tab:latency} reports these quantities measured on an NVIDIA A40. Notably, the
recurrent encoder step costs only $0.35$\,ms, making the temporal encoder a negligible
fixed per-step cost. MeanFlow completes a prediction step in $6.8$\,ms, whereas diffusion
requires $19.8$\,ms with $3$ NFE and up to $324.7$\,ms with $50$ NFE, corresponding to
$2.9\times$ to $47.5\times$ higher latency. Because the encoder cost is negligible, the
latency ratio is largely determined by $N_{\mathrm{NFE}}$ and is therefore largely
independent of the hardware. The same ordering is observed in terms of FLOPs.

\begin{table}[t]
\centering
\caption{Inference complexity.}
\label{tab:latency}
\setlength{\tabcolsep}{6pt}
\begin{tabular}{lccc}
\toprule
Model (NFE) & latency (ms)  & FLOPs (M) \\
\midrule
MeanFlow ($1$)   & $\mathbf{6.8}$  & $\mathbf{561}$ \\
Diffusion ($3$)  & $19.8$  & $917$ \\
Diffusion ($10$) & $65.2$  & $2162$ \\
Diffusion ($20$) & $130.1$  & $3941$ \\
Diffusion ($50$) & $324.7$ & $9278$ \\
\bottomrule
\end{tabular}
\end{table}

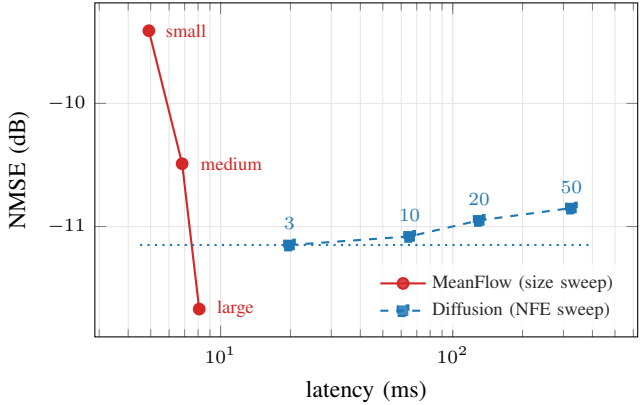
\begin{figure}[t]
\centering
\begin{tikzpicture}
\begin{axis}[width=\columnwidth,height=6.0cm,
  xmode=log, log basis x=10,
  xlabel={latency (ms)},
  ylabel={NMSE (dB)},
  grid=both, grid style={gray!20},
  legend style={font=\scriptsize,at={(0.98,0.02)},anchor=south east,draw=none,fill=white,fill opacity=0.85},
  legend cell align=left, tick label style={font=\scriptsize}, label style={font=\small}]
\addplot[cmfmu,mark=*,thick] coordinates{(4.91,-9.41)(6.83,-10.49)(8.09,-11.67)};
\addplot[cdfmu,mark=square*,thick,dashed] coordinates{(19.81,-11.15)(65.22,-11.08)(130.08,-10.95)(324.69,-10.85)};
\addplot[cdfmu,dotted,thick,forget plot] coordinates{(4.5,-11.15)(400,-11.15)};
\node[cmfmu,font=\scriptsize,anchor=west] at (axis cs:5.3,-9.41) {small};
\node[cmfmu,font=\scriptsize,anchor=west] at (axis cs:7.5,-10.49) {medium};
\node[cmfmu,font=\scriptsize,anchor=west] at (axis cs:8.8,-11.67) {large};
\node[cdfmu,font=\scriptsize,anchor=south] at (axis cs:19.81,-11.10) {$3$};
\node[cdfmu,font=\scriptsize,anchor=south] at (axis cs:65.22,-11.03) {$10$};
\node[cdfmu,font=\scriptsize,anchor=south] at (axis cs:130.08,-10.90) {$20$};
\node[cdfmu,font=\scriptsize,anchor=south] at (axis cs:324.69,-10.80) {$50$};
\legend{MeanFlow (size sweep), Diffusion (NFE sweep)}
\end{axis}
\end{tikzpicture}
\vspace{0mm}
\caption{Accuracy versus per-step latency.}
\label{fig:pareto}
\end{figure}

Finally, because the per-step latency is dominated by the generator evaluations, MeanFlow can scale its generator for higher accuracy at almost no
latency cost. Fig.~\ref{fig:pareto} plots NMSE against per-step latency for MeanFlow across
generator sizes and for diffusion across NFE (at the default backbone).
The three MeanFlow variants scale only the shared U-Net generator---small, medium, and large use base widths of $24$, $32$, and $48$ channels, respectively, resulting in $1.7$, $2.2$, and $3.8$\,M total parameters. The medium model is the default used elsewhere in the paper.
The MeanFlow frontier lies entirely below and to the left of diffusion's: increasing the
generator size drives NMSE down along a nearly vertical, low-latency line. In particular, the largest MeanFlow model attains a lower NMSE
($-11.67$\,dB) than the best diffusion configuration ($-11.15$\,dB) at roughly
half the per-step latency ($8.1$ vs.\ $19.8$\,ms). Thus, for CSI prediction, spending
compute on a larger one-step generator is far more effective than spending it on additional sampling steps for diffusion.

\section{Conclusion}
This paper proposed a one-step generative CSI prediction framework based on MeanFlow to reduce the inference complexity and latency associated with generative models in real-time wireless applications. To alleviate the difficulty of generating future CSI from pure random noise, we introduced a channel-informed prior that initializes the generation process from an informative estimate of the future channel, substantially improving both point prediction accuracy and uncertainty calibration.
We conducted evaluation using metrics that assess both point prediction accuracy and predictive distribution quality, and compared the proposed MeanFlow framework with state-of-the-art diffusion-based CSI prediction models under a controlled setting with identical encoder and generator architectures. The results show that one-step MeanFlow achieves point prediction accuracy comparable to few-step diffusion while providing substantially better-calibrated predictive uncertainty and downstream performance. Moreover, MeanFlow offers a favorable accuracy--latency scaling: increasing the one-step generator size improves NMSE with only a minor latency increase, highlighting the advantage of investing compute in a larger one-step generator rather than in additional diffusion sampling steps.

As future work, we will investigate channel-shaped source priors that exploit the structured sparsity of wireless residuals in the angle–delay domain to further improve the quality of CSI generation and uncertainty estimation.

\bibliographystyle{IEEEtran}
\bibliography{Refs}

\end{document}